\documentclass[12pt]{emulateapj} 
\usepackage{amsmath}
\usepackage{amssymb}
\usepackage{amstext}
\usepackage{graphicx}
\usepackage{epstopdf}
\usepackage{epsfig}
\usepackage{verbatim}
\usepackage{color}
\definecolor{myColor}{rgb}{0.9,0.9,0.9}    
\begin{document}
\renewcommand\bottomfraction{.9}
\shorttitle{\uppercase{Madhusudhan \& Seager}}
\title{High metallicity and non-equilibrium chemistry in the dayside atmosphere of hot-Neptune GJ~436\lowercase{b}}
\author{N. Madhusudhan\altaffilmark{1}$^{,}$\altaffilmark{2} and 
  S. Seager\altaffilmark{2}$^{,}$\altaffilmark{3}}
\altaffiltext{1}{Corresponding author:{\tt nmadhu@mit.edu}} 
\altaffiltext{2}{MIT Kavli Institute for Astrophysics and Space Research, and 
  Department of Earth, Atmospheric, and Planetary Sciences, MIT, Cambridge, MA, 02139}
\altaffiltext{3}{Department of Physics, MIT, Cambridge, MA 02139} 

\begin{abstract}

 We present a detailed analysis of the dayside atmosphere of the hot-Neptune GJ~436b, 
 based on recent {\it Spitzer} observations. We report statistical constraints on the thermal and 
 chemical properties of the planetary atmosphere, study correlations between the various 
 molecular species, and discuss scenarios of equilibrium and non-equilibrium chemistry in 
 GJ~436b. We model the atmosphere with a one-dimensional line-by-line radiative transfer code with 
 parameterized molecular abundances and temperature structure. We explore the model parameter 
 space with $10^6$ models, using a Markov chain Monte Carlo scheme. Our results encompass previous findings, indicating a paucity of methane, an overabundance of CO and CO$_2$, and a slight 
 underabundance of H$_2$O, as compared to equilibrium chemistry with solar metallicity. 
 The concentrations of the species are highly correlated. Our best-fit, and most plausible, constraints 
 require a CH$_4$ mixing ratio of $10^{-7} to 10^{-6}$, with CO $\geq 10^{-3}$, CO$_2$ $\sim 10^{-6} to 10^{-4}$, and H$_2$O $\leq 10^{-4}$; higher CH$_4$ would require much higher CO and CO$_2$. Based on calculations of equilibrium and non-equilibrium chemistry, we find that the observed abundances 
can potentially be explained by a combination of high metallicity ($\sim 10~\times$ solar) and vertical 
mixing with $K_{zz} \sim 10^6 - 10^7$ cm$^2$/s. The inferred metallicity is enhanced over that of the host star which is known to be consistent with solar metallicity. Our constraints rule out a dayside thermal inversion in GJ~436b. We emphasize that the constraints reported in this work depend crucially on the observations in the two {\it Spitzer} channels at 3.6 $\micron$ and 4.5 $\micron$. Future observations with warm {\it Spitzer} and with the {\it James Webb Space Telescope} will be extremely important to improve upon the present constraints on the abundances of carbon species in the dayside atmosphere of GJ~436b. 

\end{abstract}

\keywords{planetary systems --- planets and satellites: general --- 
planets and satellites: individual (GJ~436b) --- radiative transfer}

\section{Introduction}
\label{sec:intro}

 The last decade in exoplanetary science has demonstrated our capability 
 in detecting and characterizing atmospheres of transiting extrasolar giant 
 planets. Several observations have been reported using the {\it Hubble Space 
 Telescope (HST)}, the {\it Spitzer Space Telescope (Spitzer)}, and from ground. 
 Beginning with the first detection of sodium in the atmosphere of 
 HD~209458b in transit (Charbonneau et al. 2002), and the first detections 
 of dayside thermal emission from hot Jupiters TrES-1 and HD~209458b (Charbonneau et al. 2005; 
 Deming et al. 2005), atmospheric observations of giant exoplanets
 today are a norm, albeit still very challenging (e.g. Knutson et al. 2008; Charbonneau 
 et al. 2008; Swain et al. 2008; Grillmair et al. 2008; Desert et al. 2009; Swain et 
 al. 2009a). The intensity of observational efforts have been matched with 
 equally challenging accomplishments in theoretical modeling of exoplanet atmospheres, 
 and data interpretation (Seager \& Sasselov, 2000; Seager et al. 2005; Burrows et 
 al. 2006 \& 2008; Fortney et al. 2006; Barman et al. 2005; Tinetti et al. 2007;
 Showman et al. 2009; Madhusudhan \& Seager, 2009). Several inferences of gaseous 
 H$_2$O, CH$_4$, CO and CO$_2$, and thermal inversions have subsequently 
 been made in hot Jupiter atmospheres (Burrows et al. 2007; Tinetti et al. 2007; 
 Barman, 2007; Grillmair et al. 2008; Swain et al. 2008; 
 Madhusudhan \& Seager, 2009; Swain et al. 2009a \& 2009b). 

 A new era in exoplanetary science has now dawned. Latest observations are 
 leading to discovery and characterization of transiting exoplanets much less massive
 than the archetypal hot Jupiters, namely extrasolar Neptunes and super-Earths. 
 Several low-mass transiting planets are presently known, e.g. hot Neptunes 
 GJ~436b (Butler et al. 2004; Maness et al. 2007), HAT-P-11b (Bakos et al. 2010), 
 HAT-P-26b (Hartman et al. 2010), and Kepler-4b (Borucki et al. 2010), and super-Earths 
 CoRoT-7b (Leger et al. 2009) and GJ~1214b (Charbonneau et al.  2009). These 
 planets have opened a new regime in atmospheric modeling and data interpretation. 
 However, observations of thermal emission have been reported for only one of these 
 planets so far, GJ~436b (Deming et al. 2007, Stevenson et al. 2010). 

 The first transiting hot Neptune known, GJ~436b, with a mass of 22.6~M$_\oplus$ and radius 
 of 4.2~R$_\oplus$, orbits an M Dwarf at an orbital separation of 0.03 AU (Butler et al. 2004; 
 Gillon et al. 2007; Maness et al. 2007). The host star has an effective temperature of about 
 3500 K and a metallicity consistent with solar (Torres et al. 2008). The average density of the 
 planet is 1.5 g/cc (Torres et al. 2008), i.e., similar to the bulk density of Neptune, an ice giant. 
 At an equilibrium temperature of $\sim$ 700 K, assuming zero albedo and efficient energy circulation, 
 the density originally hinted of a hot-ice interior (Gillon et al. 2007). More detailed studies of the 
 possible bulk composition of GJ~436b indicate that an additional layer of H/He atmosphere would 
 be needed to account for the observed radius (Figueira et al.~2009; Rogers \& Seager.~2009).

The atmosphere of GJ~436b has been a subject of substantial interest in the recent past. 
Deming et al. (2007) and Demory et al. (2007) reported independent detections of thermal 
emission from the dayside of GJ~436b in the {\it Spitzer} 8 $\micron$ IRAC channel. Although 
no meaningful inferences about the molecular compositions can be drawn from a single data-point, 
model fits to the 8 $\micron$ flux contrast have favored the interpretation of inefficient day-night 
energy redistribution for GJ~436b (Demory et al. 2007; Spiegel et al. 2010, unless there are 
additional unknown optical absorbers in the atmosphere). The models used in Demory et al. 2007 
and Spiegel et al. 2010 had assumed equilibrium chemistry. Given the low temperatures of GJ~436b 
compared to hot Jupiters, equilibrium chemistry suggests that the planet's atmosphere must be abundant in methane and water vapor, and be scarce in carbon monoxide (e.g Burrows and Sharp, 1999). 

Recent observations have suggested distinct departures from predictions of equilibrium chemistry 
models. Pont et al. (2009) reported a transmission spectrum of GJ~436b obtained in the 1.1-1.9 
$\micron$ bandpass using the {\it HST} NICMOS instrument, but found no significant feature in the 
1.4 $\micron$ water band; the spectrum was flat at the 2-$\sigma$ uncertainties. More recently, 
Stevenson et al. (2010) reported planet-star flux contrasts of the dayside atmosphere of GJ~436b 
in six channels of {\it Spitzer} broadband photometry, and inferred a deficiency of methane 
in the atmosphere of GJ~436b, using models based on Madhusudhan \& Seager (2009) and 
the present work. The high planet-star flux contrast observed in the 3.6 $\micron$ IRAC 
channel was central to the low methane requirement. Even though our models do not assume 
equilibrium chemistry, the inferred methane mixing ratio of $\sim 10^{-7}$ in a hydrogen dominated 
atmosphere at $\sim 700 K$ signals a surprising new regime in atmospheric chemistry of extrasolar planets. 

In this work, we report detailed statistical constraints on the atmospheric properties of  GJ~436b, 
and explore channels of equilibrium and non-equilibrium chemistry that might explain the observed 
chemical abundances. We first estimate the atmospheric chemical composition and temperature 
structure at different levels of fit to the data, using a 1-D line-by-line radiative transfer model for 
exoplanet atmospheres (Madhusudhan \& Seager, 2009). Our constraints result from exploring 
the model parameter space with $\sim 10^6$ models, optimized using a Markov chain Monte 
Carlo optimization scheme. We then use detailed calculations of equilibrium and non-equilibrium 
chemistry, and with different metallicities, to explain the observed constraints on the various 
molecular species. 

Our results indicate that a high metallicity and non-equilibrium chemistry are required to explain 
the molecular abundances constrained by the observations. The requirement of non-equilibrium 
chemistry is consistent with our findings in Stevenson et al. (2010). The observations require the 
presence of CO and CO$_2$ and a substantial depletion of CH$_4$ in the dayside atmosphere of 
GJ~436b. In this work, we find that the constraints on the CO and CO$_2$ mixing ratios can be 
explained by vertical mixing in the atmosphere (with $K_{zz} = 10^6 - 10^7$ cm$^2$/s) and a 
high metallicity ($10 \times $ solar). And, we suggest that the low CH$_4$ required can potentially 
be explained by non-equilibrium chemistry. We also find that the observations require inefficient 
day-night energy redistribution in GJ~436b, which can be confirmed by future observations of 
phase curves. 

In what follows, we first give a brief overview of model independent interpretation of {\it Spitzer} 
observations in section~\ref{sec:data_interpretation}. We then describe our atmosphere model, 
the parameter space optimization scheme, and the models for equilibrium and non-equilibrium 
chemistry we use in this work, in section~\ref{sec:model}. We present the results of our analysis 
in section~\ref{sec:results}. Finally, in section~\ref{sec:discussion}, we present a summary of our 
work and discuss consequences for future observations and theoretical models.

\begin{figure}
\centering
\includegraphics[width = 0.5 \textwidth]{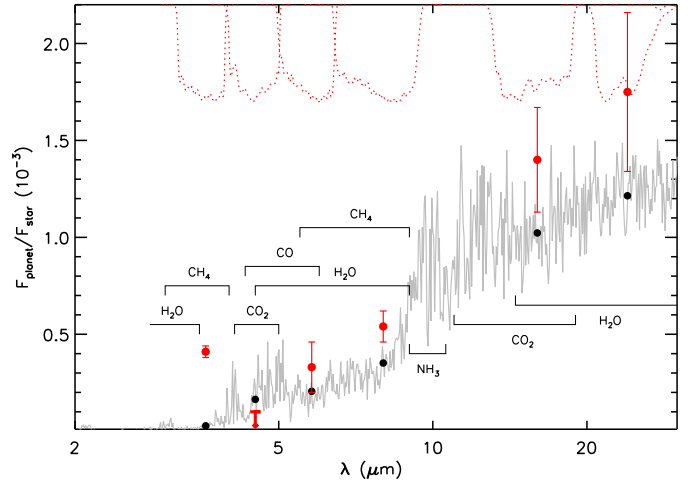}
\caption{ Molecular absorption features in {\it Spitzer} photometric bandpasses. 
The red dotted lines at the top show the six {\it Spitzer} bandpasses. The black lines 
show the extent of absorption features due to the corresponding molecules. 
The gray curve shows a hypothetical model spectrum of GJ~436b based on equilibrium 
chemistry, and the black filled circles show the corresponding integrated points in the 
{\it Spitzer} channels. The red filled circles with error bars show the observations of GJ~436b reported by Stevenson et al. (2010).}
\label{fig:spitzer}
\end{figure}

\section{Overview of Spitzer Data Interpretation}
\label{sec:data_interpretation}

The concentrations of major molecular species in an exoplanetary atmosphere can 
be constrained to some extent from observations in the six channels of {\it Spitzer} 
photometry, at 3.6 $\micron$, 4.5 $\micron$, 5.8 $\micron$, 8 $\micron$, 16 $\micron$, and 
24 $\micron$. Figure~\ref{fig:spitzer} illustrates the spectral features of the dominant molecules 
in the {\it Spitzer} channels, for a synthetic thermal spectrum of GJ 436b. CH$_4$ has strong 
features almost exclusively in the 3.6 $\micron$ and 8 $\micron$ channels. CO has a strong feature 
in the 4.5 $\micron$ channel, also contributing to the 5.8 $\micron$ channel. And, CO$_2$ has 
strong features in the 4.5 $\micron$ and 16 $\micron$ channels. Although H$_2$O has several 
spectral features in most of the {\it Spitzer} channels, the stronger and/or unique features lie 
in the 3.6 $\micron$, 5.8 $\micron$ and 24 $\micron$ channels. Given the strong features of the 
molecules in one or more {\it Spitzer} channels, reasonable constraints can be placed with high 
S/N observations. In principle, however, the presence of additional species in significant amounts, 
due to non-equilibrium chemistry (e.g. Zahnle et al. 2009b), and non-LTE effects (Swain et al. 2010) 
can also affect the emergent spectrum. We discuss these factors in section~\ref{sec:model-model}. 
 
The constraints on the atmospheric properties, that might be possible from high S/N {\it Spitzer} photometry, 
can be understood in the light of the key parameters effecting the emergent spectrum (Madhusudhan \& Seager, 2009). 
Under the assumption that H$_2$O, CH$_4$, CO and CO$_2$, are the dominant spectroscopically active 
molecules in the {\it Spitzer} bands, the molecular concentrations constitute four free parameters. And, 
although the temperature structure can involve many free parameters, the temperature at the base of the 
atmosphere (at P $\sim$ 1 - 10 bar), and the thermal gradient, are the two most important 
parameters, in the absence of a thermal inversion. Apart from the data themselves, an additional constraint 
on the parameters appear in the form of energy balance (discussed in section~\ref{sec:model-model}). Thus, the 
six observations can potentially lead to meaningful constraints on the six key atmospheric parameters, 
mainly on the four dominant molecules. These same molecular species can then provide constraints on the 
C/H and O/H ratios in the atmosphere. 

As an example of constraining molecular abundances from the {\it Spitzer} data, consider a planetary 
atmosphere with T$_{\rm eq} \sim 700 - 1000$ K, and a temperature profile decreasing outward 
(i.e., no thermal inversion). Further suppose that a low thermal flux is observed in the 3.6 $\micron$ channel. 
Such a low flux would suggest strong absorption due to a high concentration of methane. And, if the 
low 3.6 $\micron$ flux is indeed due to a high methane abundance (since water vapor also absorbs weakly in
the 3.6 $\micron$ channel), the flux should also be low in the 8 $\micron$  channel where methane also 
absorbs strongly (as shown, for example, by the black filled circles in Figure~\ref{fig:spitzer}). On the other 
hand, high  planet fluxes in the 3.6 and 8 $\micron$ channels indicate low absorption due to methane, 
and hence a paucity of methane in the planet atmosphere. 

A second example concerns constraining the molecular abundances of CO and CO$_2$. A low flux 
observed in the 4.5 $\micron$ channel must indicate atmospheric absorption due to CO and/or CO$_2$. 
At 4.5 $\micron$, CO$_2$ has stronger absorption cross-section than CO, so that a small concentration 
of CO$_2$ can produce an absorption feature comparable to that from a relatively large concentration of  CO. 
The degeneracy between CO and CO$_2$ contributions in a 4.5 $\micron$ measurement can be broken 
by an observation in the 16 $\micron$ channel, where only CO$_2$ contributes, among the two. As a third 
example, constraints on H$_2$O are based primarily on fluxes in the 5.8 $\micron$ and 24 $\micron$ 
channels. An important note concerns the presence of  a thermal inversion in the atmosphere, a region 
where temperature increases outward. In such a case, the molecular features would be emission features 
instead of absorption features, thus reversing the logic of inferences described above (Madhusudhan \& Seager, 2010). 

The recent observations of the dayside atmosphere of GJ~436b by 
Stevenson et al.~(2010) represent a quintessential example of the above 
inferences. Their observations indicate an extremely high flux in the 3.6 
$\micron$ channel and an extremely low flux in the 4.5 $\micron$ channel, 
causing a brightness temperature differential of $\sim$ 450 K between the 
two adjacent channels. Thus, following the arguments described above, the 
observations point towards an extremely low methane abundance and high 
CO and/or CO$_2$ abundances, for a temperature profile without  a thermal 
inversion. This identification of low CH$_4$ and high CO and/or CO$_2$ was in 
fact the central result of Stevenson et al. (2010), based on a more elaborate 
atmospheric modeling procedure, also discussed in the present work. We emphasize 
that, although the six channels of photometry can yield statistical constraints on 
the atmospheric properties, they cannot yield a unique solution, given the large 
number of free parameters (Madhusudhan \& Seager, 2009). 

\section{Model}
\label{sec:model}
 
  Our goal is to determine the best fitting interpretation for observations of the 
  dayside atmosphere of GJ~436b. We first fit the data with a large ensemble 
  of 1-D dayside atmosphere models of GJ~436b, and determine regions of the 
  parameter space that fit  the data best. Our results yield best-fit constraints on 
  the molecular abundances and temperature structure. We then use some of 
  the best-fit temperature profiles, along with independent calculations of 
  equilibrium and non-equilibrium chemistry to see if we can explain the 
  observed constraints on the molecular abundances. 
  
 \subsection{Radiative Transfer Model}
 \label{sec:model-model}
   
 In order to fit the observations with model spectra, we use the 1-D exoplanetary 
 atmosphere model developed in Madhusudhan \& Seager (2009). Our model 
 consists of a line-by-line radiative transfer code, with constraints of hydrostatic 
 equilibrium and global energy balance, and coupled to a parametric pressure-temperature 
 (P-T) structure and parametric molecular abundances. This modeling approach allows 
 one to compute large ensembles of models, and to efficiently explore the parameter 
 space of molecular abundances and temperature structure. 
 
 The major difference of our model from traditional 1-D atmosphere models is in the 
 treatment of energy balance. Our model requires energy balance at the top of the 
 atmosphere, instead of an iterative scheme to ensure layer-by-layer radiative 
 (or radiative + convective) equilibrium which is assumed in conventional models 
 (e.g. Seager et al.~2005, Burrows et al.~2006, Fortney et al.~2006).  
  For a given set of model parameters, we require that the net energy output at the top of 
  the atmosphere is less than or equal to the net energy input due to the incident 
  stellar flux; a deficit indicates energy redistributed to the night-side. Models where 
  the emergent flux exceeds the incident flux are discarded (see Madhusudhan \&
 Seager, 2009).  
 
 In this work, we consider well mixed atmospheres, i.e. uniform mixing ratio of each 
 molecular species over the entire atmosphere. The present approach allows us to 
 sample a wider range of compositions independent of any assumptions of equilibrium 
 chemistry. The molecular species in our models include molecular hydrogen (H$_2$), water 
 vapor (H$_2$O), carbon monoxide (CO), carbon dioxide (CO$_2$), and methane (CH$_4$). 
 We have used an alternate approach in previous works (e.g. Madhusudhan \& Seager, 2009), 
 where we parameterized the abundances in terms of deviations from equilibrium chemistry. 
 We find that the constraints on the overall mixing ratios do not depend critically on the 
 choice of parametrization. Our H$_2$O, CH$_4$, and CO molecular line data are from 
 Freedman et al. 2008, and references therein. Our CO$_2$ data are from Freedman 
 (personal communication, 2009) and Rothman et al. (2005). And, we obtain the H$_2$-H$_2$ 
 collision-induced opacities from Borysow et al. (1997), and Borysow (2002). 

 Our model does not include effects of non-LTE radiative transfer. Models of 
  exoplanetary spectra with Non-LTE radiative transfer calculations have 
  not been reported. Recently, Swain et al. (2010) reported detection of excess 
  emission at 3.25 $\micron$  in a ground-based thermal spectrum  of HD~189733b, 
  which they surmised to be due to non-LTE methane emission. However, follow-up 
  observations of Mandell et al. (2010) failed to detect the feature reported 
  by Swain et al. (2010). Furthermore, their estimates of potential contribution from resonant 
  florescence, a non-LTE emission mechanism, of methane yielded fluxes too low to 
  significantly contribute to the emission spectrum, contrary to the observations of 
  Swain et al. (2010). Nevertheless, detailed atmospheric models of GJ~436b in the 
  future might need to account for non-LTE contributions to spectra, as high-resolution 
  observations become possible in the future with the {\it James Webb Space Telescope}. 
  
 We also do not include in our list of molecules higher hydrocarbons that might be potential 
  byproducts of non-equilibrium chemistry. Non-equilibrium chemistry can lead to hydrocarbons like 
  C$_2$H$_2$ and C$_2$H$_4$, depending on the temperature and degree of vertical 
  mixing (Zahnle et al. 2009b; Line et al. 2010). However, as will be discussed in 
  section~\ref{sec:photochem}, the observations of GJ~436b indicate high temperatures 
  ($T \gtrsim 1100$K)  at 1-10 bar pressures, thereby favoring the oxidation of CH$_4$ to 
  CO, over conversion of CH$_4$ to higher hydrocarbons (Zahnle et al. 2009b). Nevertheless, 
  constraining these species observationally would still be a worthwhile exercise when higher 
  resolution observations become available with the {\it JWST}

\begin{figure}
\centering
\includegraphics[width = 0.5 \textwidth]{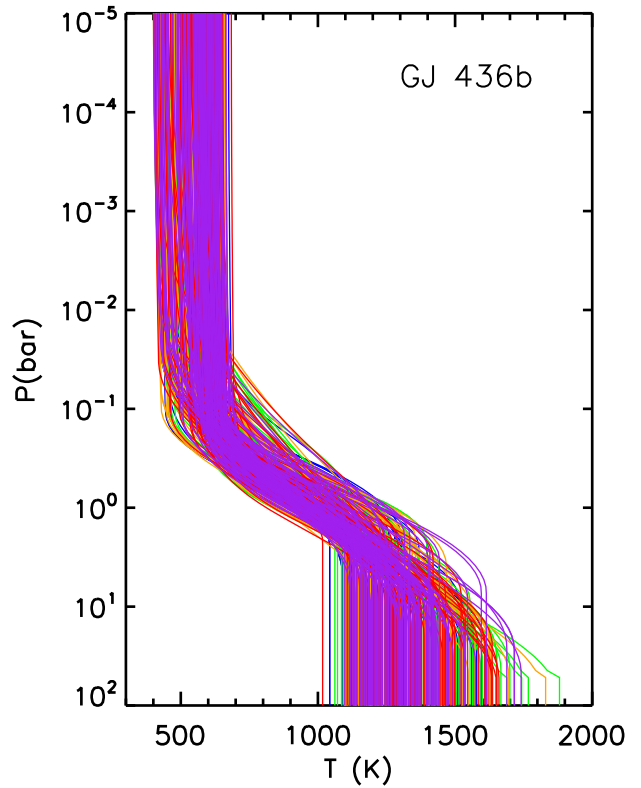}
\caption{Pressure-Temperature (P-T) profiles of GJ 436b.The purple, 
red, orange, green and blue profiles correspond to models which fit the observations to 
within $\xi^2$ of 1.0, 2.0, 3.0, 4.0 and 5.0, respectively.}
\label{fig:pt}
\end{figure}

\begin{figure*}[ht]
\centering
\includegraphics[width = \textwidth]{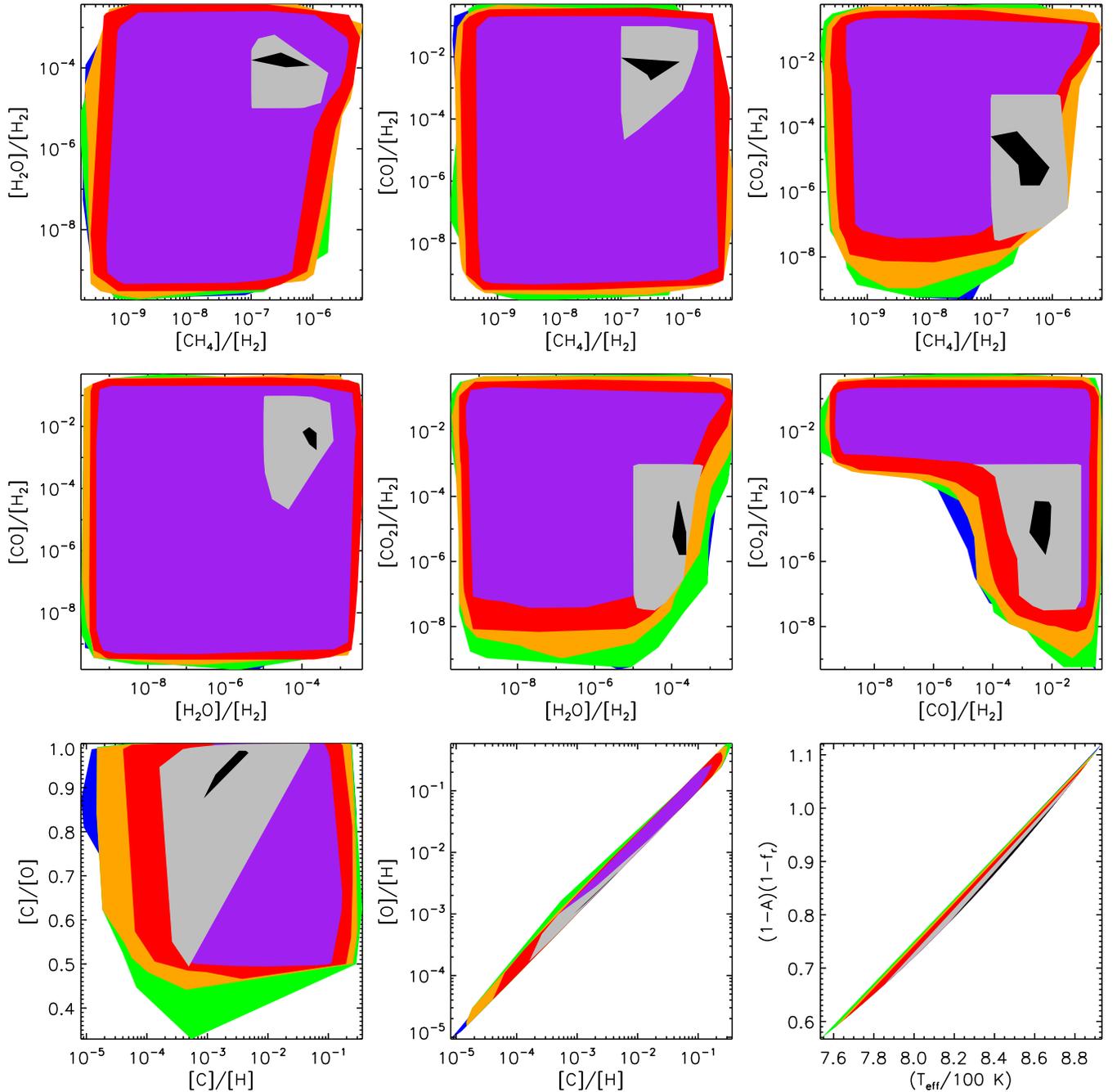}
\caption{Constraints on the atmospheric properties of GJ 436b. Mixing ratios are shown as ratios by 
number density. The contours show surfaces of minimum $\xi^2$ in the space of atmospheric composition 
and temperature structure. The purple, red, orange, green and blue, correspond to minimum 
$\xi^2$ of 1.0, 2.0, 3.0, 4.0 and 5.0, respectively. The gray and black regions correspond to models 
which have methane mixing ratios greater than $10^{-7}$, along with different conditions on the 
compositions of the remaining molecules, and allowing fits within $\xi^2 = 3$. For the gray 
surfaces, CO$_2 \leq 10^{-3}$, H$_2$O $\geq 10^{-5}$, and CO $ \leq 10^{-1}$. CO$_2$ 
of $\sim 10^{-3}$ and  CO of $\sim10^{-1}$ are implausible either in equilibrium or non-equilibrium 
chemistry (Lodders \& Fegley, 2002; Zahnle et al. 2009a), however, we show these solutions 
for completeness. The black surfaces show solutions within more plausible limits 
(see section~\ref{sec:plausibility}) of CO$_2 \leq 10^{-4}$, H$_2$O $\geq 10^{-4}$, and 
CO$ \leq 10^{-2}$. The black contour in the C/H - O/H plane requires C/H and O/H $\geq10 \times$ 
solar abundances.}
\label{fig:cplot}
\end{figure*}

\begin{figure*}
\centering
\includegraphics[width = 6in]{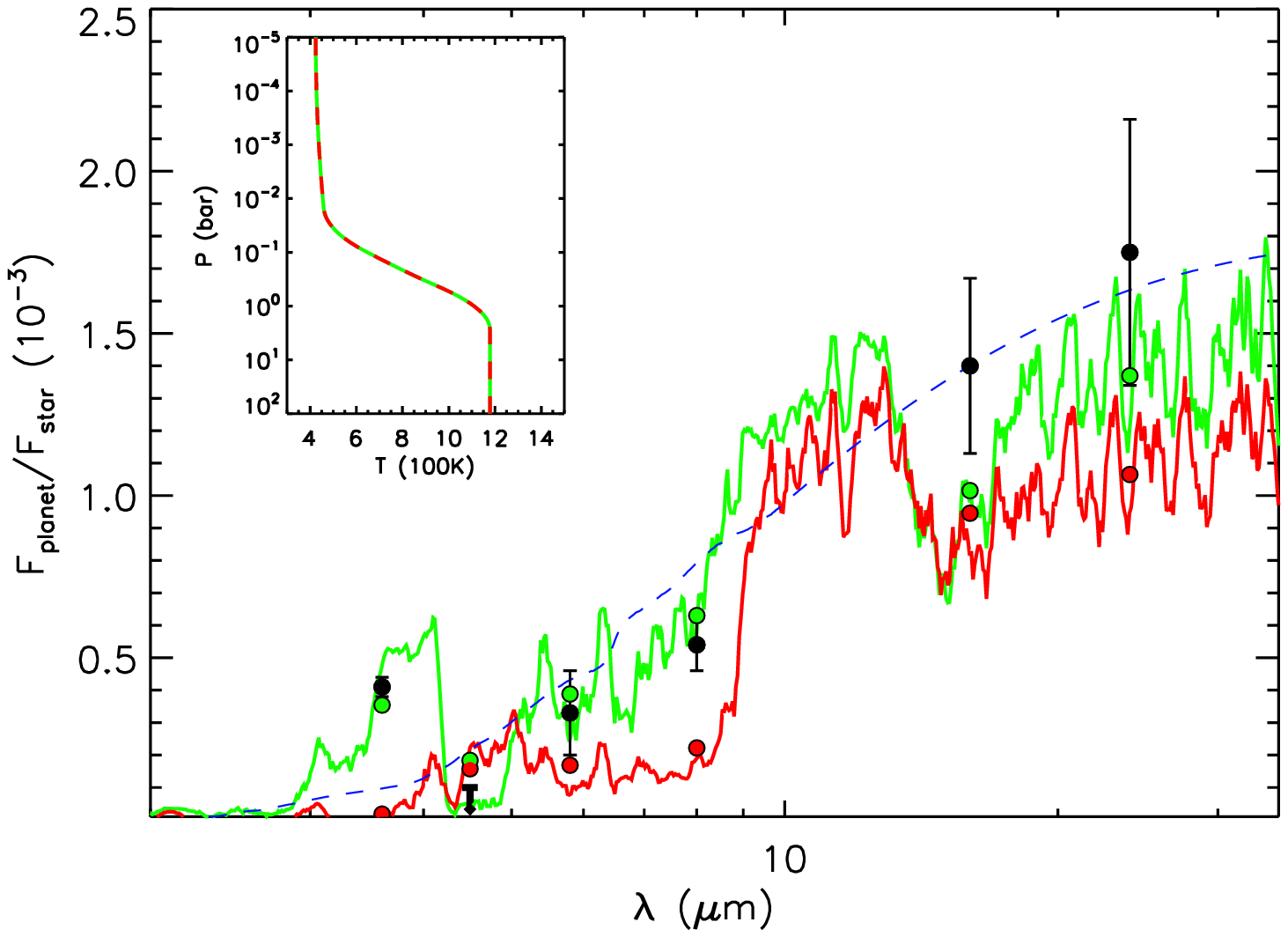}
\caption{Observations and model spectra for dayside emission from GJ 436b. The black filled circles with 
error bars show the {\it Spitzer} observations in the six photometric channels, from Stevenson et al. 2010. 
The blue dashed lines shows a planet blackbody spectra at 800 K. The green and red curves 
show two model spectra, and the colored circles show the corresponding channel integrated 
model points. The green model is a best-fit model spectrum (see section~\ref{sec:plausibility}, with 
non-equilibrium molecular mixing ratios of H$_2$O $= 10^{-4}$ , CO =  $7 \times 10^{-3}$, CH$_4$ $= 10^{-6}$, 
and CO$_2$ =  $ 6 \times 10^{-6}$. The red model has a composition close to chemical equilibrium with 
solar abundances, with H$_2$O $= 2 \times 10^{-3}$ , CO =  $10^{-5}$, and CH$_4$ $= 7 \times 10^{-4}$; it also contains CO$_2$ $= 10^{-6}$. Both models have the same pressure-temperature profile, shown in the inset. 
The green, non-equilibrium, model has a maximum day-night energy redistribution fraction ($f_r$) of 0.03, i.e for zero bond albedo ($A_B$). On the other hand, the red, equilibrium, model has very efficient redistribution, 
$f_r \leq 0.63$, for $A_B = 0$.} 

\label{fig:spectra}
\end{figure*}

\subsection{Parameter Space Exploration for Model Fits}
\label{sec:model-mcmc}
We use the Markov chain Monte Carlo (MCMC) method to explore the 
model parameter space. The MCMC method is a Bayesian parameter 
estimation algorithm which allows the calculation of posterior probability 
distributions of the model parameters conditional to a given set of 
observations, and prior probabilities (see e.g. Tegmark et al. 2004; 
Ford, 2005). In this work, our goal is not parameter estimation -- the 
number of model parameters ($N = 10$) exceeds the number of available 
observations ($N_{\rm obs} = 6$) and rendering the problem under-constrained . 
However, the MCMC method allows an efficient means of exploring the 
parameter space in search of regions which provide the best fits to the 
observations. We, therefore, use the MCMC method with a Metropolis-Hastings
scheme within the Gibbs sampler, for fine sampling of the model parameter space . 
And, we report error contours in the space of the molecular compositions 
and temperature structure. Our statistic of choice is $\xi^2$, defined as 
$\chi^2/N_{\rm obs}$ (Madhusudhan \& Seager, 2009). In this metric, a 
$\xi^2 = 1$ indicates models fitting the observations within the 1-$\sigma$ 
observational uncertainities, on average. Similarly, $\xi^2$ of 2 and 3 indicate 
fits at the 1.41-$\sigma$  (i.e $\sqrt{2}$) and 1.73 -$\sigma$ error bars, respectively.

Our model described in \S~\ref{sec:model-model} above has ten free parameters 
(Madhusudhan \& Seager, 2009). Six parameters concern the $P$-$T$ profile: T$_0$, P$_1$, 
P$_2$, P$_3$, $\alpha_1$, and $\alpha_2$ . And, four parameters correspond to 
the uniform molecular mixing ratios:  $f_{\rm H_2O}$, $f_{\rm CO}$, $f_{\rm CH_4}$, 
and $f_{\rm CO_2}$. We define the mixing ratio of a molecule as the number fraction 
with respect to molecular hydrogen. 

We define some physically motivated boundaries in the parameter space explored 
by the Markov chain. We restrict all the molecular mixing ratios to a range 
of $[10^{-10}, 0.1]$. We also impose the constraint of global energy balance by 
restricting $\eta$ to [0.0,1.0], where, $\eta = (1-A)(1-f_r)$ is the ratio of emergent flux 
output on the dayside to incident stellar flux input on the dayside, weighted appropriately 
(Madhusudhan \& Seager, 2009). Here, $A$ is the Bond Albedo and $f_r$ is the day-night 
energy redistribution. The ``fit'' parameters for the MCMC are T$_0$, $\log$(P$_1$), 
$\log$(P$_2$), $\log$(P$_3$), $\alpha_1$, $\alpha_2$, $\log$($f_{\rm H_2O}$), 
$\log$($f_{\rm CO}$), $\log$($f_{\rm CH_4}$), and $\log$($f_{\rm CO_2}$). We consider 
uniform priors in all the parameters. For each system under consideration, we run one chain 
of $10^6$ links, which takes $\sim 22$ hours on a single processor.

 \subsection{Chemistry Model}
\label{sec:model-chemistry}
After we obtain the constraints on the molecular abundances from model fits to 
data, we investigate processes of atmospheric chemistry that could explain 
the required abundances. Atmospheres of giant planets in the solar system and 
those of brown dwarfs have revealed the interplay between equilibrium and non-equilibrium 
chemical processes in hydrogen-rich atmospheres (Prinn \& Bashay, 1977; Fegley \& Lodders, 
1994; Noll et al., 1997; Saumon et al. 2006). At high pressures deep in a planetary atmosphere, 
molecules react fast enough that all species are in thermochemical equilibrium. As the pressures 
decrease with increasing altitude, thermochemical reaction rates decrease, allowing 
for competing non-equilibrium processes with shorter timescales to shift the involved 
species out of equilibrium. We compute the atmospheric compositions in equilibrium 
using the equilibrium chemistry code adapted from Seager et al. (2005).

\subsubsection{Equilibrium chemistry}
\label{sec:equib}

Our objective here is to determine whether the observed constraints 
on the molecular mixing ratios are consistent with chemical equilibrium. 
To this end, we calculate the equilibrium compositions of the species 
using the equilibrium chemistry code originally developed in 
Seager et al. (2000), and subsequently used in Seager et al. (2005) and 
Miller-Ricci et al. (2009). We calculate the gas phase molecular mixing ratios for 172 molecules, resulting from 
abundances of 23 atomic species, by minimizing the net Gibbs free energy of the 
system. The multi-dimensional Newton-Raphson method described in White et al. (1958) 
was used for the minimization. We adopt polynomial fits for the 
free energies of the molecules, based on Sharp \& Huebner (1990). 
We assume a hydrogen dominated atmosphere for GJ~436b, and we compute 
equilibrium concentrations of all the species at varying metallicites 
(see section \ref{sec:results-chemical}), over a grid in pressure and 
temperature. 

At the temperatures of GJ~436b, the most abundant and spectroscopically 
active molecules in the {\it Spitzer} bandpasses are expected to be H$_2$O, CH$_4$, 
CO and CO$_2$. Ammonia (NH$_3$) should also be abundant, but does not contain major 
features in the {\it Spitzer} channels. At high temperatures, e.g. $\gtrsim 1300$ K 
at 1 bar pressure, like those in hot Jupiter atmospheres, CO is predicted to be 
the dominant carbon and oxygen bearing species. At lower temperatures, on the 
other hand, CH$_4$ is the dominant carbon bearing species in equilibrium. Water 
vapor is a major carrier of oxygen in either regime. The specific amounts of 
each of these species also depend strongly on the assumed metallicity, and 
pressure. Finally, the amount of carbon dioxide in equilibrium is a very strong 
function of metallicity. At the temperatures of GJ~436b, solar metallicity 
yields a CO$_2$ mixing ratio up to $10^{-7}$, whereas amounts as high as 
$10^{-4}$ can be obtained for 30 $\times$ solar metallicity.

\subsubsection{Non-equilibrium chemistry}
\label{sec:nonequib}

Vertical mixing can drive species out of equilibrium in regions of the atmosphere where 
the timescale of vertical transport is shorter compared to the timescale governing chemical 
equilibrium between the relevant species. At high pressures, deep in the atmosphere, 
convection is a natural mixing mechanism. However, above the radiative-convective boundary, 
atmospheric instabilities and turbulent processes, such as wave breaking, can also lead to vertical 
motions. This form of mixing in the radiative zone is collectively termed as `eddy mixing' or `eddy diffusion', 
and is parameterized as a diffusion process, with a coefficient of eddy diffusion ($K_{zz}$).  
Such mixing shifts molecular species, in radiative regions, away from their equilibrium 
concentrations. Eddy mixing has been known in the context of atmospheres of solar system giant 
planets (Prinn \& Bashay, 1977; Fegley and Lodders, 1994), and  brown dwarfs (Noll et al. 1997; 
Saumon et al. 2003 \& 2006; Hubeny \& Burrows, 2007). 
And, Cooper \& Showman (2006) studied eddy mixing in hot Jupiter atmospheres.  
Eddy mixing offers a viable explanation to the excess carbon monoxide discovered in hydrogen 
dominated atmospheres (see e.g Yung \& Demore, 1999). The primary reaction governing the relative 
abundances of CH$_4$ and CO in equilibrium is given by: 

\begin{equation}
   {\rm CO + 3H_2  \rightleftharpoons CH_4 + H_2O}.
   \label{eq:eddy_1}
\end{equation}

This reaction favors CO at high temperatures and CH$_4$ at the low temperatures of GJ~436b. Thus, 
based on equilibrium chemistry, it is expected for CO to be dominant in hotter lower regions of the 
atmosphere ($T \gtrsim 1200 K$), and CH$_4$ to be dominant at higher altitudes where temperatures 
are lower ($T \lesssim 1000 K$). However, atmospheric spectra of solar system planets and cooler brown 
dwarfs suggest significant amounts of CO in the upper layers of the atmosphere 
(see e.g. Fegley and Lodders, 1994; Noll et al. 1997, Stephens et al. 2009). This is achieved by eddy 
mixing which vertically transport CO from the lower regions of an atmosphere to the upper regions. 

Eddy mixing dominates when the mixing time scale ($\tau_{\rm mix}$) is shorter than the chemical 
time scale ($\tau_{\rm chem}$) of CO in the forward reaction in (\ref{eq:eddy_1}). The forward 
reaction in (\ref{eq:eddy_1}) in fact proceeds via multiple steps, and Yung et al. (1988) suggested 
the rate determining step in the reaction chain to be (but, cf. Visscher et al. 2010): 

\begin{equation}
   {\rm H + H_2CO  + M \rightarrow CH_3O + M}
   \label{eq:eddy_2}
\end{equation}
The life time of CO is given by: 

\begin{equation}
  \tau_{\rm chem} \sim \frac{\rm [CO]}{\rm d[CO]/dt} = \frac{\rm [CO]}{k_f {\rm [H][H_2CO]}}, 
   \label{eq:eddy_3}
\end{equation}
where, $k_f$ is the rate constant for the forward reaction. $k_f$ is not known directly, but can be 
estimated from the reverse reaction rate constant ($k_r$), which is known from laboratory experiments, 
and  the equilibrium constant ($K_{eq}$) for the reaction (see Griffith \& Yelle, 1999, for a detailed discussion). 
In the present study, we use the following estimate of $k_f$, based on Line et al. (2010):
\begin{equation}
  k_f = 3.07 \times 10^{-12} T^{-1.2}e^{(3927/T)}
   \label{eq:eddy_4}
\end{equation}
Then, $\tau_{\rm chem}$ can be calculated from (\ref{eq:eddy_3}), using equilibrium 
concentrations of the CO, H and H$_2$CO. 

The mixing time ($\tau_{\rm mix}$) in radiative regions of the atmosphere 
is determined by the eddy diffusion coefficient ($K_{zz}$) and a characteristic length scale 
for mixing ($L$), as 
\begin{equation}
  \tau_{\rm mix} \sim \frac{L^2}{K_{zz}} 
   \label{eq:eddy_5}
\end{equation}

$L$ is typically chosen to be the scale height ($H$) (e.g. Prinn \& Bashay, 1977; Line et al. 2010). However, for the coolest of giant planets, like Jupiter, $L$ can be as low as 0.1$H$ (Smith, 1998). In this work, we choose $L = H$; a lower $L$ does not significantly alter our results, as discussed in section~\ref{sec:co_co2}. 

As is evident from the above discussion, both $\tau_{\rm mix}$ and $\tau_{\rm chem}$ vary with height 
in the atmosphere; although, $\tau_{\rm mix}$ varies to a lesser extent than
$\tau_{\rm chem}$. But, while $\tau_{\rm chem}$ increases towards higher levels in the atmosphere (i.e 
with decreasing pressure), $\tau_{\rm mix}$ increases in the opposite direction. The pressure ($p_0$) at which 
$\tau_{\rm mix} = \tau_{\rm chem}$ is called the ``quench" level. Above this pressure, i.e in deeper layers 
of the atmosphere, the species are in chemical equilibrium, and below this pressure the concentration 
of the species (CO in this case) is fixed, or ``quenched", at the equilibrium value at $p_0$. This yields a 
uniform mixing ratio profile for the species for pressures below $p_0$.

Thus, a higher $p_0$ implies that CO can be dredged up from deeper levels in the atmosphere. And, 
since in equilibrium CO concentration increases with pressure, a high $p_0$ implies higher concentration 
of CO in the upper layers of the atmosphere.  It can be shown that, for a given $\tau_{\rm chem}$ profile, 
$p_0$ increases monotonically with $K_{zz}$, thus correlating a high CO concentration in the upper 
atmosphere with a high $K_{zz}$ (see for e.g. Griffith \& Yelle, 1999). In the discussion here, we have 
assumed a temperature profile that increases monotonically with pressure. The dependence of observed 
CO on $K_{zz}$ deviates from this monotonic behavior for more complicated temperature structures, for 
example in the presence of thermal inversions or partial isotherms, as will be shown in 
section~\ref{sec:results-nonequib}. 

In this work, we explore the CO mixing ratios resulting from different combinations of 
$K_{zz}$ and metallicities. We explore values of $K_{zz}$ between 10$^2$ - 10$^{10}$, 
and metallicities of solar -- 30 $\times$ solar. Our choices of metallicities are motivated by 
the constraints on CO and CO$_2$ abundances which indicate high C/H and O/H ratios 
apriori. And, our range in $K_{zz}$ encompasses values found in solar system planets 
and brown dwarfs. For comparison, planetary atmospheres in the solar system have 
$K_{zz}$ ranging between $10^5 - 10^9$ cm$^2$/s, and $K_{zz}$ for brown dwarf 
atmospheres can be as low as $10^2 -10^4$ cm$^2$/s (Prinn and Bashay, 1977; 
Yung and Demore, 1999; Saumon et al. 2003). 

\begin{figure*}[ht]
\centering
\includegraphics[width = \textwidth]{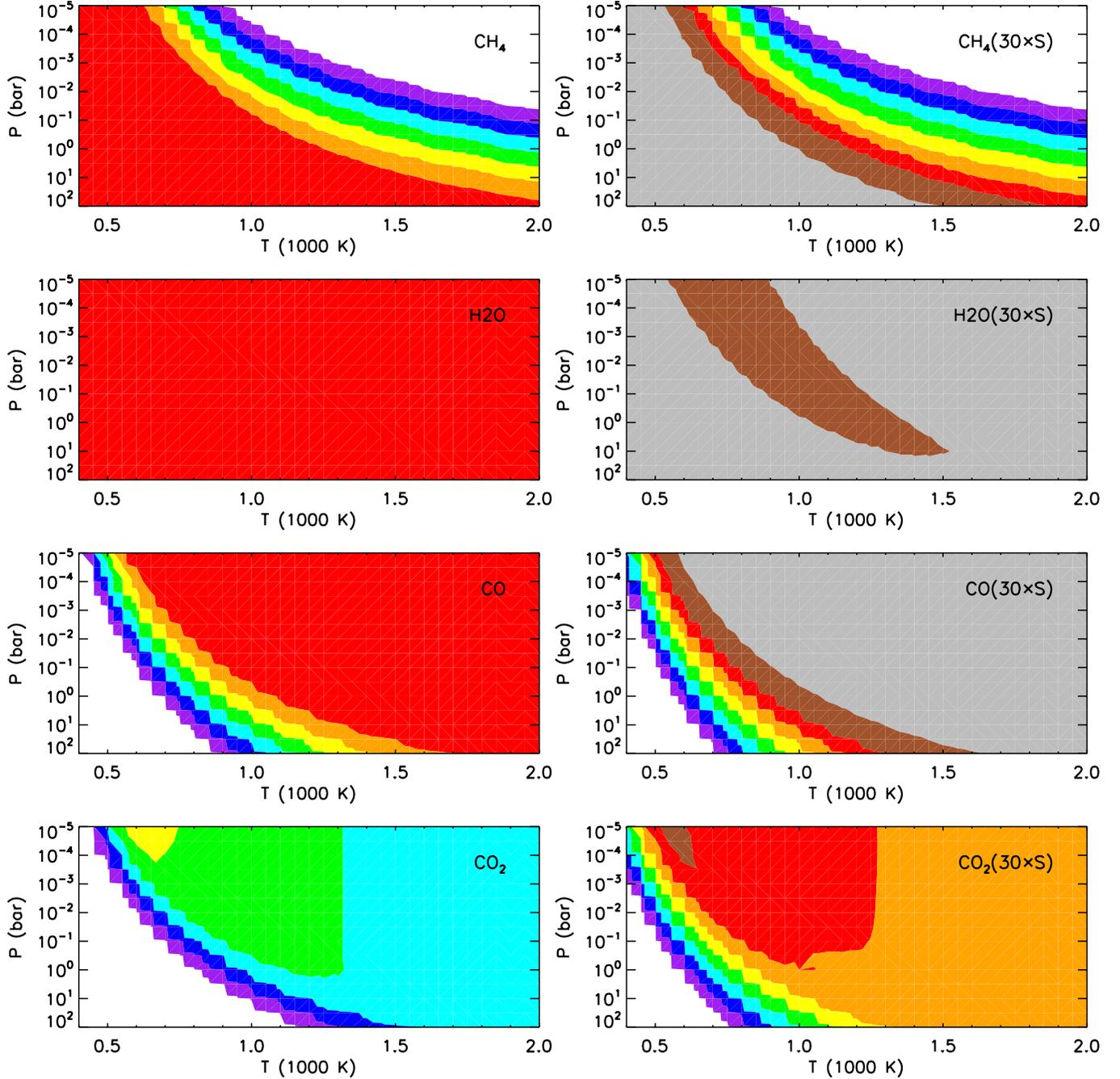}
\caption{Concentrations of CH$_4$, H$_2$O, CO, and CO$_2$, predicted by equilibrium chemistry assuming solar and 30$\times$solar elemental abundances (see section \ref{sec:equib}). Each panel shows contours of molecular mixing ratios (i.e. ratio by number density) with respect to molecular hydrogen, in pressure-temperature space. The left panels show mixing ratios for solar metallicity, and the right panels (``30$\times$S") have 30 $\times$ solar metallicity. The gray, brown, red, orange, yellow, green, cyan, blue and purple contours correspond to mixing ratios greater than $10^{-2}$, $10^{-3}$, $10^{-4}$, $10^{-5}$, $10^{-6}$, $10^{-7}$, $10^{-8}$, $10^{-9}$, and $10^{-10}$, respectively. The black lines are some best fitting pressure-temperature profiles. }
\label{fig:equib}
\end{figure*}

\section{Results}
\label{sec:results}

In this section, we present constraints on the atmospheric properties of the dayside atmosphere 
of GJ 436b, as placed by the six channel {\it Spitzer} photometry.  We first report constraints on 
the molecular abundances, and discuss the correlations between the various species. We then 
discuss the physical plausibility of the solutions, and present calculations of equilibrium and 
non-equilibrium chemistry attempting to explain the observed constraints. Finally, we present 
constraints on the atmospheric temperature structure and day-night energy redistribution. Our 
constraints follow from a detailed exploration of the model parameter space with $\sim 10^6$ models, 
using the procedures described in section~\ref{sec:model-mcmc}. 

Our modeling approach allows placing statistically robust model constraints from the data. The pressure-temperature ($P$-$T$) profiles explored by our models are shown in Figure~\ref{fig:pt}, color coded by their degree of fit to data. 
The goodness-of-fit contours in the space of atmospheric composition are shown in Figure~\ref{fig:cplot}. 
 As mentioned in section~\ref{sec:model-mcmc}, our statistic for the goodness-of-fit is given by (Madhusudhan \& 
 Seager, 2009):
\begin{equation}
\xi^2 = \frac{1}{N_{obs}} \sum_{i = 1}^{N_{obs}} \bigg(\frac{f_{i,model} - f_{i,obs}}
{\sigma_{i,obs}} \bigg)^2,
\label{eq:xi}
\end{equation}
where, $f_{i,model}$ is the planet-star flux contrast of the model in each channel, 
and $f_{i,obs}$ and $\sigma_{i,obs}$ are the observed flux contrast and the 1-$\sigma$ 
uncertainty, in that channel. $N_{obs}$ is the number of observations. Here, $N_{obs} = 6$, 
corresponding to the six channels of {\it Spitzer} photometry. In this metric, a 
$\xi^2 = 1$ indicates models fitting the observations within the 1-$\sigma$ 
observational uncertainities, on average. Similarly, $\xi^2$ of 2 and 3 indicate 
fits at the 1.41-$\sigma$  (i.e $\sqrt{2}$) and 1.73 -$\sigma$ error bars, respectively.

The constraints depend on the $\xi^2$ surface one chooses for interpretation, apart from 
any conditions of physical plausibility one would like to impose on the models.  We first 
report constraints at the $\xi^2 = 1$ and $\xi^2 = 2$ levels, and with only the barest assumptions 
of physical plausibility. We then discuss additional constraints that result from considering some 
nominal conditions of physical plausibility, from equilibrium and non-equilibrium chemistry. 

\subsection{Constraints on Chemical Composition}
\label{sec:results-chemical}

The constraints on the molecular abundances are strongly influenced by the correlations between 
them. The correlations between the molecules result from their overlapping absorption features in 
the {\it Spitzer} channels as described in section \ref{sec:data_interpretation}. We present constraints 
on the mixing ratios of methane (CH$_4$), water vapor (H$_2$O), carbon monoxide (CO), and carbon 
dioxide (CO$_2$). Because the abundances of molecules are correlated, constraints on any molecule 
have to be discussed with respect to abundances of one or more of the remaining molecules. 

The constraints on all the molecules and the correlations between them are shown in Figure~\ref{fig:cplot}. 

{\it Methane} CH$_4$: Our results indicate a substantial paucity of methane in the 
dayside atmosphere of GJ 436b. Our results place an absolute upper-limit on the mixing 
ratio of methane to be $3\times10^{-6} - 6\times10^{-6}$, for $\xi^2$ ranging between 1 -- 5, 
and assuming nothing about the remaining molecules. However, these upper-limits allow 
for a wide range of abundances of the remaining molecules, including some manifestly 
impractical values. Primarily, the constraints include CO$_2$ abundances as high 0.3, 
implying 30\% of a hydrogen dominated atmosphere to be composed of CO$_2$! 
Assuming a high metallicity for the planet atmosphere (about 30 $\times$ solar), CO$_2$ 
mixing ratios as high as $\sim 10^{-4}$ can be attained by equilibrium chemistry, as shown 
in Figure~\ref{fig:equib} (also see Zahnle et al. 2009a, 2009b). 

The methane mixing ratio is constrained to values below 10$^{-6}$, if we impose a plausible 
limits on the CO$_2$ abundance. A generous upper-limit on the CO$_2$ abundance can be 
assumed to be $\sim 10^{-3}$, based on the arguments above. Allowing a maximum 
CO$_2$ of $10^{-3}$, at the $\xi^2 \leq 1$ surface (purple surfaces in Figure~\ref{fig:cplot}) the 
methane mixing ratio is 
constrained to between $10^{-7} - 10^{-6}$, for CO$_2$ mixing ratios between $10^{-7} - 10^{-3}$. 
And, at the $\xi^2 \leq 2$ surface (red surfaces in Figure~\ref{fig:cplot}), CH$_4$ = $10^{-7} - 10^{-6}$, 
for CO$_2$ $ = 10^{-8} - 10^{-3}$. There is no lower bound on the CH$_4$ abundance; mixing 
ratios below $10^{-9}$ do not have discernible features at the resolution of the current data. 

The low methane requirement is enforced primarily by the hight planet-star flux contrast in the 
3.6 $\micron$ {\it Spitzer} IRAC channel. Inflating the uncertainties in the 3.6 $\micron$ channel 
does not obviate the low methane requirement. And, the strong correlation of methane with CO$_2$ 
arises from the large flux differential between the 3.6 $\micron$ and 4.5 $\micron$ channels, 
as has been described in section~\ref{sec:data_interpretation}. Methane is also correlated 
with water vapor which also has features in the 3.6 $\micron$ channel, albeit to a lesser extent, 
as shown in Figure~\ref{fig:cplot}. 

The low mixing ratio of methane is a clear indication of non-equilibrium chemistry in the dayside atmosphere of GJ~436b, as has been suggested in Stevenson et al. (2010). Equilibrium chemistry at the temperatures of GJ~436b causes methane to be dominant carbon bearing molecule. At solar abundances the methane mixing ratio in chemical equilibrium, for typical temperature profiles of GJ~436b, is predicted to be 
$7 \times 10^{-4}$, and $2 \times 10^{-2}$ for 30 $\times$ solar abundances, as is evident from Figure~\ref{fig:equib}. 

{\it Water vapor} (H$_2$O): Our results place an absolute upper-limit on the H$_2$O abundance, 
as shown in Figure~\ref{fig:cplot}. 
The H$_2$O mixing ratio is constrained to $< 10^{-3}$ and $< 3 \times 10^{-3}$ for $\xi^2 \leq 1$ 
and $\xi^2 \leq 5$, respectively, if we make no assumptions of physical plausibility of the solutions. 
As in the case of methane, however, the H$_2$O abundance is also correlated with the CO$_2$ 
abundance. If we restrict CO$_2$ to a generous upper-limit of $10^{-3}$, as described for the case 
of methane, the $\xi^2 \leq 1$ and $\xi^2 \leq 2$ surfaces constrain the H$_2$O abundance to 
$< 3 \times 10^{-4}$ and $< 10^{-3}$, respectively. H$_2$O is also correlated with CO and CH$_4$.
Thus the H$_2$O abundance can be further constrained if we assume conditions of physical 
plausibility of all the species simultaneously (discussed below in section~\ref{sec:plausibility}). 
The correlations of H$_2$O with all the remaining molecules arise from it numerous features in 
all the {\it Spitzer} channels, as described in section~\ref{sec:data_interpretation}.

{\it Carbon monoxide} (CO): The abundance of CO is highly correlated with the abundance 
of CO$_2$. If no assumption is made on the CO$_2$ abundance, the observations 
provide no constraint on the CO mixing ratio, as shown in Figure~\ref{fig:cplot}. If the mixing ratio of CO$_2$ is 
restricted to the conservative upper-limit of $10^{-3}$, the observations constrain 
CO to be $ \geq 3 \times 10^{-5}$ for $\xi^2 < 1$, while still leaving it unconstrained at  
$\xi^2 \sim 2$ and higher. 

Despite the constraints on CO above, simultaneous conditions of physical plausibility on all 
the molecules indicate that a very high CO abundance ($\geq 10^{-3}$) is essential to explain the 
observations. This will be discussed in section~\ref{sec:plausibility} below. The strong correlation 
between CO and CO$_2$ arises from the fact that both molecules have 
strong features in the 4.5 $\micron$ IRAC channel, the CO$_2$ feature being stronger, 
as described in section~\ref{sec:data_interpretation}. Although not apparent in Figure~\ref{fig:cplot}, 
CO is also correlated with H$_2$O and CH$_4$, via their correlations with CO$_2$.

{\it Carbon dioxide} (CO$_2$): The observations require a clear presence of CO$_2$ in the 
atmosphere, as shown in Figure~\ref{fig:cplot}. Since CO$_2$ and CO are correlated, a low 
concentration of CO$_2$ requires a high concentration of CO. At the $\xi^2 \leq 1$ surface, 
a CO$_2$ concentration less than $10^{-7}$ requires a CO concentration greater than $10^{-2}$. 
On the other hand, having a CO$_2$ concentration of $10^{-4}$ allows for CO concentrations 
as low as $\sim 10^{-3}$, at the $\xi^2 \leq 1$ surface. For $\xi^2$ surfaces of 2 and higher, 
lower CO abundances can fit the data for a given CO$_2$ abundance, as shown in Figure~\ref{fig:cplot}. 
However, CO$_2$ is also correlated with H$_2$O, such that an H$_2$O abundance greater than 3 $\times 10^{-4}$ would require a CO$_2$ abundance greater than $\sim 10^{-4}$, at the $\xi^2 \leq 1$ surface. 

The observations themselves do not place any upper-limit on the CO$_2$ abundance. However, the 
maximum amount of CO$_2$ possible can be constrained based on theoretical limits of 
equilibrium and non-equilibrium chemistry. 
For the best-fit temperature profiles of GJ 436b shown in Figure~\ref{fig:pt}, and pressures in the $10^{-3} - 1$ bar range, thermochemical equilibrium can yield CO$_2$ mixing ratios up to 10$^{-7}$ for solar metallicity and 
up to 10$^{-4}$ for $\sim 30~\times$ solar metallicities (see Figure~\ref{fig:equib}). The strong dependence 
of CO$_2$ abundance on metallicity has been reported before (Lodders \& Fegley, 2002; Zahnle et al. 2009a)

\subsubsection{Plausibility of the Abundance Constraints}
\label{sec:plausibility}

The constraints reported above assume nothing with regards to the physical plausibility of 
the models, except the conservative limit of CO$_2$ $\leq 10^{-3}$ used for sake of argument. 
However, reasonable theoretical constraints can be placed over the observed constraints based on 
well established arguments of equilibrium and non-equilibrium chemistry (see section~\ref{sec:model-chemistry}). The mixing ratios of CH$_4$, H$_2$O, CO and CO$_2$ under chemical equilibrium are shown in 
Figure~\ref{fig:equib}, for a range of temperatures and pressures pertinent to GJ~436b, along 
with some best-fit $P$-$T$ profiles. A detailed discussion of non-equilibrium chemistry, via eddy mixing, 
is presented in section~\ref{sec:results-nonequib}. 

We find that stringent constraints on the molecular abundances required by the data can be 
placed even with modest assumptions of atmospheric chemistry. As alluded to in 
section~\ref{sec:results-chemical} above, and shown in Figure~\ref{fig:equib}, CO$_2$ 
mixing ratios up to $10^{-4}$ are allowed for high metallicity (also see Zahnle et al. 2009a; Lodders \& Fegley, 2002). H$_2$O is a major carrier of oxygen in the desired temperature range. The H$_2$O mixing 
ratio is expected to be $\sim 10^{-4}$ and $\sim 5 \times 10^{-3}$, for solar and 30$\times$ solar abundances, 
respectively.  And, while there is no lower limit on the CO abundance, the CO upper-limit is fixed 
by the metallicity; CO $\lesssim 10^{-4}$ and $\lesssim 10^{-2}$, for solar and 30$\times$solar abundances, 
respectively. Finally, the concentration of methane in equilibrium follows the carbon abundance. At the 
temperatures of GJ~436b, methane is supposed to be highly abundant in equilibrium, as shown in 
Figure~\ref{fig:equib}, with mixing ratios of $7 \times 10^{-4}$, and $2 \times 10^{-2}$ for solar and 
30 $\times$ solar abundances, respectively. However, CO can be enhanced and methane can be 
depleted to some extent due to non-equilibrium chemistry (Zahnle et al. 2009b), which will be discussed 
in section \ref{sec:results-nonequib} below.

The constraints due to the considerations of physical plausibility are shown in Figure~\ref{fig:cplot}. 
The gray surface shows regions assuming conservative boundaries of $\xi^2 \leq 3$, 
CH$_4$ $\geq 10^{-7}$, CO$_2 \leq 10^{-3}$, H$_2$O $\geq 10^{-5}$, and CO$ \leq 10^{-1}$. 
And, the black surfaces show a subset of the gray surface with $\xi^2 \leq 3$, CH$_4$ $\geq 10^{-7}$, 
CO$_2 \leq 10^{-4}$, H$_2$O $\geq 10^{-4}$, and CO $ \leq 10^{-2}$. The black contours represent 
our most likely interpretation of the observations, which will be justified below. A best-fit model consistent 
with our constraints above is shown in Figure~\ref{fig:spectra}.

Our most plausible constraints on the atmosphere of GJ~436b indicate the possibility of high metallicity, 
along with non-equilibrium chemistry. The black surfaces in Figure~\ref{fig:cplot} show that a CO abundance 
$\geq 10^{-3}$ is required to have a H$_2$O abundance of $\geq 10^{-4}$ and a methane 
abundance above $10^{-7}$. The corresponding constraint on CO$_2$ is $10^{-6} - 10^{-4}$. 
While this CO$_2$ abundance can be explained based on equilibrium chemistry with 
high metallicity alone (see Figure~\ref{fig:equib}), the high CO abundance requires a high 
metallicity along with non-equilibrium chemistry.  As shown in Figure~\ref{fig:equib}, 
very high CO abundances can exist at the bottom of the atmosphere, for high metallicity, however the cooler upper layers of the atmosphere have much lower CO abundance.  Non-equilibrium chemistry in the form of eddy 
mixing can transport CO from the lower layers to the upper layers of the atmosphere to cause  a uniformly high CO over the entire atmosphere, as discussed in section \ref{sec:nonequib}. On the other hand, the low CH$_4$ abundance could potentially be caused by non-equilibrium chemistry as well (Zahnle et al. 2009b). In what follows, we will explore the realm of non-equilibrium chemistry in an attempt to explain the observed constraints. 

\begin{figure*}[ht]
\centering
\includegraphics[width = \textwidth]{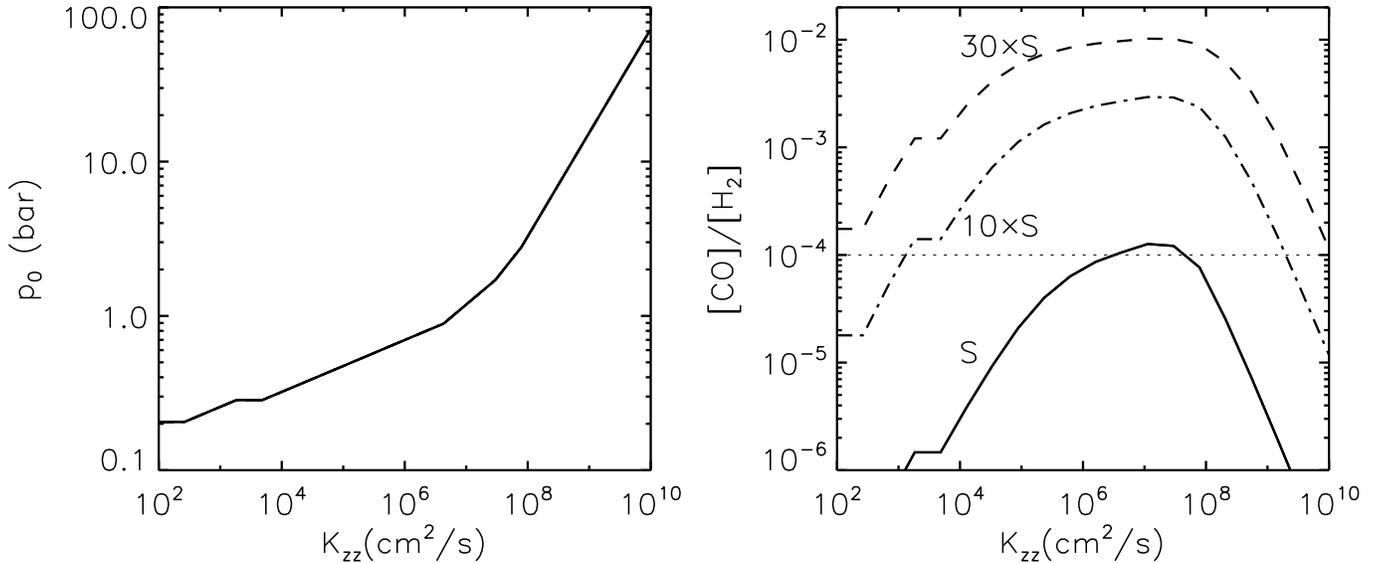}
\caption{Vertical eddy diffusion in GJ~436b. The two panels show the dependence of the ``quench" pressure 
(p$_0$) and the CO mixing ratio on the coefficient of eddy diffusion, $K_{zz}$.
``S", $10 \times S$, and $30 \times S$ refer to solar, 10 times solar and 30 times solar metallicities.}
\label{fig:eddy}
\end{figure*}

\begin{figure*}[ht]
\centering
\includegraphics[width = \textwidth]{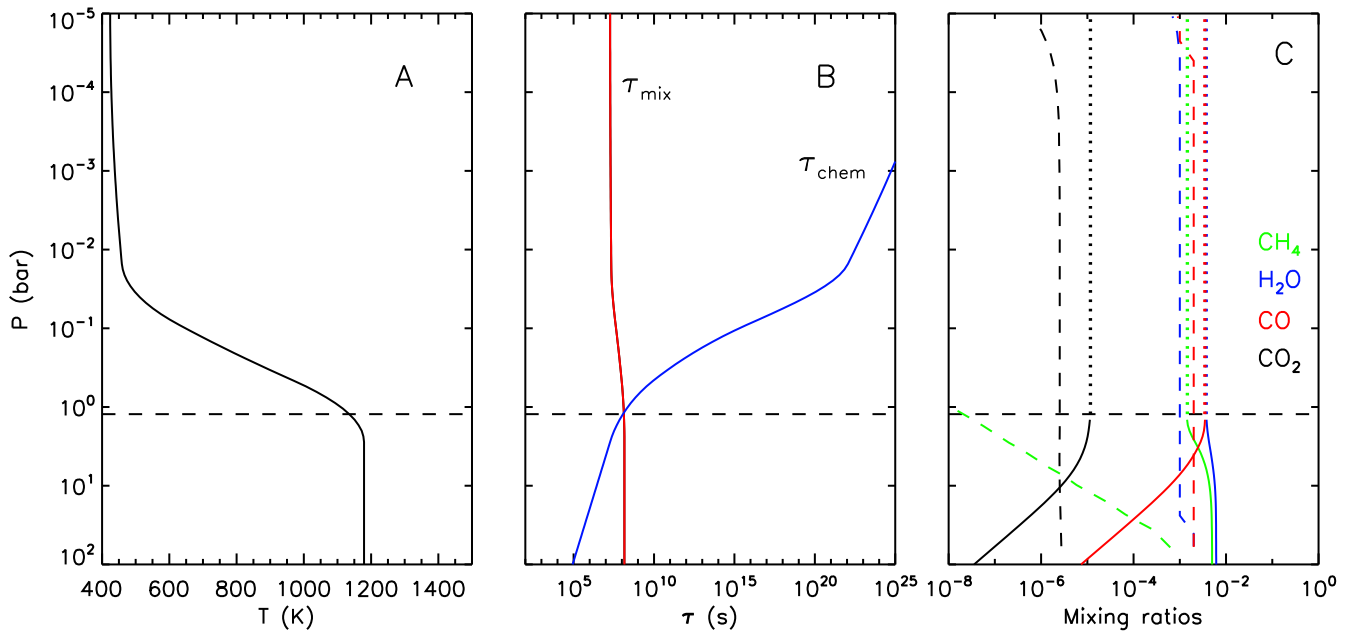}
\caption{Non-equilibrium chemistry in GJ~436b. Panel A shows the pressure-temperature 
($P$-$T$) profile of a best-fit model. Panel B shows the times scales of eddy mixing 
($\tau_{\rm mix}$) and of CO--CH$_4$ equilibrium chemistry ($\tau_{\rm chem}$). 
Panel C shows the influence of eddy mixing on the mixing ratios of the four prominent species. 
The solid lines in panel C show the compositions from equilibrium chemistry, which continue as dotted lines above the quench level, and fixed to the values at the quench level. The dashed lines show the 
mixing ratios from Zahnle et al. (2009b), for an isothermal atmosphere with $T = 1200K$, $K_{zz} = 10^7$ cm$^2$/s, and $5\times$solar metallicity, which yields a CH$_4$ mixing ratio below 10$^{-6}$ at the $P = 1$ bar level. The y-axis in all the panels is pressure in bars. The horizontal black dashed line in all the panels  show 
the quenching pressure level.}
\label{fig:photo}
\end{figure*}

\subsection{Explanations for Non-equilibrium Abundances}
\label{sec:results-nonequib}

Our best-fit constraints on the chemical composition of GJ~436b require substantial 
deviations from equilibrium chemistry with solar metallicity. Non-equilibrium processes 
have been known to influence chemical compositions of several planetary and brown 
dwarf atmospheres, as discussed in section~\ref{sec:model-chemistry}. Here, we explore 
channels of non-equilibrium chemistry in search of potential explanations to the observed 
constraints which are inexplicable by equilibrium chemistry alone - namely the high CO and CO$_2$ 
abundances, and the low CH$_4$ abundance. 

\subsubsection{High CO and CO$_2$}
\label{sec:co_co2}
The high abundances of CO and CO$_2$ required by the observations 
can be achieved via eddy mixing, along with a high metallicity. As explained in section~\ref{sec:nonequib}, 
eddy mixing transports CO from the deeper layers of the atmosphere to the upper layers of the atmosphere where CO is less abundant. The relevant quantity is the quench pressure (p$_0$), which 
denotes the pressure level in the atmosphere below which the rate of vertical mixing is faster 
than the reaction rate for chemical equilibrium; the CO concentration in the upper atmosphere 
($P <$ p$_0$) freezes at its value at $P = $ p$_0$. Figure~\ref{fig:eddy} shows the dependence of the quench level (p$_0$) on the diffusion coefficient ($K_{zz}$), for a best-fit P-T profile of GJ~436b. Higher values of $K_{zz}$ lead to mixing of species from deeper levels of the atmosphere, i.e higher p$_0$. It can be seen from the 
figure that $K_{zz}$ values between 10$^6$ - 10$^7$ can cause mixing from a quench 
level of $\sim 1$ bar.  

The abundance of CO in the upper atmosphere depends on $K_{zz}$ and the metallicity. 
The right panel of Figure~\ref{fig:eddy} shows the range of CO mixing ratios that are 
attainable with different values of $K_{zz}$ and  metallicities. For solar metallicity, a 
maximum CO mixing ratio of $\sim 10^{-4}$ is attainable for $K_{zz}$ values between 
$\sim 10^7 - 10^8$.  However, CO mixing ratios above $10^{-3}$ require metallicities 
greater than $10~\times$ solar, and $K_{zz} \sim 10^5 - 10^8$. For a very high metallicity 
of 30 $\times$ solar, a maximum CO mixing ratio $10^{-4}$ can be attained for a rather 
low $K_{zz}$ of $\sim 10^3$ cm$^2$/s, and a maximum of $\sim 10^{-2}$ for $K_{zz}$ 
between $\sim 10^6 - 10^8$ cm$^2$/s. For comparison, planetary atmospheres in the 
solar system have $K_{zz}$ ranging between $10^5 - 10^9$ cm$^2$/s  (see e.g.  Prinn 
and Bashay, 1977; Fegley \& Lodders, 1994; Yung and Demore, 1999), and for 
brown dwarf atmospheres, $K_{zz}$ can range between $10^2 -10^6$ cm$^2$/s 
(Saumon et al. 2006; Stephens et al. 2009). 

The constraint on $K_{zz}$ is also dependent on the choice of the characteristic 
length scale for mixing ($L$), through equation~\ref{eq:eddy_5}. In the above results, 
we used the typical assumption of $L = H$, the scale height of the atmosphere (e.g. 
Prinn \& Bashay, 1977; Line et al. 2010). However, for the coolest of giant planets, like 
Jupiter, $L$ can be $\sim 0.1 H$ (Smith, 1998). We find that a choice of  
$L < H$ for GJ~436b results in only a modest change in the required $K_{zz}$. 
For instance, the peak CO mixing ratio for 10$\times$ solar metallicity, for the $L = H$ 
case is attained for $K_{zz} = 10^6 - 10^8$ cm$^2$/s, where as the same value for even 
the extreme case of $L = 0.1 H$ is attained for $K_{zz} = 10^4 - 10^6$ cm$^2$/s. Such 
differences in $K_{zz}$ can be compensated by uncertainty in the metallicity, as seen in Figure~\ref{fig:eddy}. 

Finally, a high metallicity is also consistent with the high CO$_2$ abundance required to explain the 
observations. At the quench levels of $\sim 1$ bar, the equilibrium mixing ratio of CO$_2$ is 
$10^{-7}$ and $10^{-4}$, for solar and 30 $\times$ solar metallicity, as shown in Figure~\ref{fig:equib}. 
The strong dependence of CO$_2$ on metallicity has been reported is previous studies (Lodders \& Fegley, 2002; Zahnle et al. 2009a), suggesting CO$_2$ as a key metallicity indicator. We, thus, find that a metallicity of 
10 - 30 $\times$ solar, and vertical mixing of $K_{zz} \sim 10^5 - 10^8$, can simultaneously explain the 
constraints on CO and CO$_2$ observed in section~\ref{sec:plausibility}. The  $K_{zz}$  and metallicity 
are constrained further in section~\ref{sec:joint}.

\subsubsection{Low CH$_4$} 
\label{sec:photochem}

The low methane abundance required by the observations cannot be explained by equilibrium 
chemistry. At $P \sim 1$ bar and $T \sim 1000$ K (corresponding to the best-fitting P-T profiles), 
equilibrium chemistry yields methane mixing ratios of about $10^{-4}$ and $10^{-2}$, for solar and 
30 $\times$ solar metallicity, respectively, as shown in Figure~\ref{fig:equib}. Furthermore, considering 
non-equilibrium thermochemistry using the CO-CH$_4$ reaction pathway alone, as investigated in section 
\ref{sec:co_co2} above, does not explain the low methane abundances. By non-equilibrium thermochemistry, we mean departures from equilibrium chemistry due to vertical mixing. The remaining alternatives in searching for a low-methane solution include a more detailed treatment of non-equilibrium thermochemistry, including 
all the possible species and reactions, and photochemistry. Zahnle et al. (2009b) have reported such 
calculations over a range of temperatures and eddy mixing coefficients that are relevant to the current 
situation. 

Substantial depletion of methane below equilibrium values is possible due to non-equilibrium 
thermochemistry and photochemistry. Full kinetics models yield a substantial amount of free radicals, especially H (Liang et al. 2003; Zahnle et al. 2009b). Zahnle et al. (2009b) 
show that the resultant overabundance of H leads to depletion of methane via two main channels. At high temperatures ($T >$ 1200 K), H readily reacts with H$_2$O to yield the reactive OH 
radical, which oxidizes CH$_4$ in very short timescales. The result is an excess production of CO. At 
lower temperatures ($T <$ 1200 K), water is more stable, leading to a reducing environment with abundant  
free H. Methane, being still reactive, reacts with H to form higher hydrocarbons, and 
can be markedly reduced below equilibrium levels. Thus, in either scenario, the methane concentration 
at the observable pressure levels can be substantially depleted below equilibrium, depending on the temperature, eddy mixing coefficient ($K_{zz}$), metallicity, and photochemistry (Zahnle et al. 2009b).

Two key parameters governing non-equilibrium methane chemistry 
are the temperature and $K_{zz}$. The best-fitting $P$-$T$ profiles required by the observations all 
have temperatures over 1100 K in the lower atmospheres ($P \gtrsim 10$ bar), similar to those obtained 
from self-consistent models of GJ~436b (Spiegel et al., 2010). Figure~\ref{fig:photo} shows a 
sample best-fit $P$-$T$ profile (panel A), which has an isothermal temperature structure of 1200 K below 
$P \sim 1$ bar. And, as described in section~\ref{sec:co_co2}, the observational constraint on 
the CO concentration requires $K_{zz}$ $\sim$ 10$^7$ cm$^2$/s. 
Panel B of Figure~\ref{fig:photo} shows the CO-CH$_4$ reaction timescale and the 
eddy mixing time scale, for $K_{zz}$ of 10$^7$ cm$^2$/s, varying with pressure; the time scales 
intersect at the quench pressure (p$_0$) of $\sim$ 1 bar. 

The methane mixing ratios attainable via kintetics and photochemistry for the required 
parameters are discussed in Zahnle et al. (2009b). The parameters are: $T = 1200$ K, $K_{zz} =$ 
10$^7$ cm$^2$/s, and p$_0$ = 1 bar. Panel C of Figure~\ref{fig:photo} shows mixing ratio profiles 
from our CO-CH$_4$ disequilibrium model (section \ref{sec:co_co2}), and the mixing ratio profiles from 
Zahnle et al. (2009b), corresponding to the same T and $K_{zz}$. As shown in Panel C, their results 
show that CH$_4$ $\leq 10^{-7}$ is possible at the 1 bar level, assuming 5 $\times$ solar metallicity 
and a stellar irradiation that  is 100 $\times$ the solar insolation at Earth. Lower temperatures and higher 
$K_{zz}$ both favor higher methane concentration at the 1 bar level. And, the chemistry is less 
sensitive to photochemistry at higher pressures deep in the atmosphere. 

\subsubsection{Joint Constraints on Metallicity and Eddy Mixing} 
\label{sec:joint}

The observational constraints on all the molecules yield a plausible set of constraints on 
the metallicity and $K_{zz}$. A full kinetic and/or photochemical model 
is beyond the scope of the current work. However, based on the models of Zahnle et al. (2009b) with the 
T and $K_{zz}$ that we constrain for GJ~436b, we find that the low methane mixing ratio ($10^{-7} - 10^{-6}$) 
observed can most likely be explained by non-equilibrium chemistry. Although such low methane abundances 
can be obtained for $5~\times$ solar metallicity and $K_{zz} \sim$ 10$^7$ cm$^2$/s, a higher metallicity would 
be required to simultaneously explain the high CO and CO$_2$ abundances, as described in section \ref{sec:co_co2}. On the other hand, too high of a metallicity, of say 30 $\times$ solar, might also increase the CH$_4$ abundance to above the favorable levels. We therefore choose an intermediate value of 10 $\times$ solar which satisfies the CO and CO$_2$ constraints. 

Based on the above reasoning, the observed chemistry in the dayside atmosphere of GJ~436b likely results from a high metallicity ($\sim 10~\times$ solar) and a $K_{zz} \sim 10^6 - 10^7$ cm$^2$/s. A more robust conclusion would be possible with a full kinetic + photochemical model of GJ~436b, using our  best-fit P-T profile, and the appropriate UV spectrum, or flux scaling, for the host star. The present constraints indicate a significant enhancement in the metallicity of GJ~436b over that of the host star whose metallicity is known to be consistent with solar (Torres et al. 2008). Additionally, the observations constrain the C/O ratio between 0.5 and 1.0, with the most likely solutions (black surfaces in Figure~\ref{fig:cplot}) suggesting a C/O between $\sim$0.85 and 1. However, whether the methane depletion required by the data can be obtained in a high C/O environment, e.g. of C/O = 1, needs to be investigated by a full non-equilibrium chemistry model in the future. 

\subsection{Temperature Structure and Day-Night Energy Redistribution}

The {\it Spitzer} observations provide important constraints on the vertical 
thermal gradient and the energy balance in the dayside atmosphere of GJ~436b. 
The observed brightness temperatures in the six {\it Spitzer} 
channels range from about 700 K in the 4.5 $\micron$ IRAC channel (3-$\sigma$ upper limit)
to about 1100 K in the 3.6 $\micron$ channel (Stevenson et al. 2010). With the exception 
of the 3.6 $\micron$ IRAC channel, observations in all the remaining five channels are 
consistent with a black-body planet spectrum at 750 K $\pm$ 100 K. However, the 3.6 $\micron$ 
observation of $\sim$ 1100 K brightness temperature is  a major exception, requiring 
a temperature differential of $\sim$ 400 K between the 3.6 $\micron$ and 4.5 $\micron$ 
channels, implying a very steep temperature gradient in the atmosphere. The dayside 
pressure-temperature (P-T) profiles of GJ 436b constrained by the observations are 
shown in Figure~\ref{fig:pt}. The best fitting P-T profiles (in purple) have temperatures varying  
by over $\sim$ 400 K per bar of pressure in the lower atmosphere, required primarily by the 
large temperature differential described above. 

The observations rule out the presence of a significant thermal inversion in the dayside 
atmosphere of GJ~436b. Our results show that the observations cannot be fit with an inversion model for any 
chemical composition (although very small inversions which might cause only weak 
observable features cannot be ruled out by present data).  A significant 
thermal inversion in this atmosphere would have caused the brightness temperatures in the 4.5, 5.8, 
and 8 $\micron$ channels to be markedly higher than the 3.6 $\micron$ channel, much higher than 
what the current data indicate. Finally, as has been known for all hot Jupiters, the data indicate 
that the observable dayside atmosphere of GJ~436b is mostly radiative, with the radiative zones of some of the 
best fitting P-T profiles extending to pressures above $\sim$ 10 bar. The isotherms at the high 
pressure ends of the P-T profiles are suggestive of the radiative diffusion approximation in the high 
optical depth limit (see e.g. Madhusudhan \& Seager, 2009), and are also found in other self-consistent 
models reported in the literature (e.g. Spiegel et al. 2010). 

The large brightness temperature observed in the 3.6 $\micron$ channel, of 1100 K, is also 
indicative of low day-night energy redistribution in GJ~436b. The 3.6 $\micron$ channel probes levels
deep in the atmosphere (around pressures of 1 bar or higher). A high 3.6 $\micron$ brightness 
temperature, therefore, indicates a high blackbody continuum emerging from the base of the dayside 
atmosphere at $\sim$ 1 bar. Our best-fit models show that the net emergent flux on the dayside nearly 
balances the incident stellar flux, implying that very little energy is circulated to the night side. The bottom-right panel of figure \ref{fig:cplot} shows the $\xi^2$ contours in the $\eta = (1-A)(1-f_r)$ and T$_{\rm eff}$. 
$\eta = (1-A)(1-f_r)$ is obtained from energy balance, where $A$ is the bond albedo and 
$f_r$ is fraction of energy redistributed to the night side (Madhusudhan \& Seager, 2009). 

The best fitting models favor $\eta \gtrsim 0.75$, or a maximum day-night 
redistribution ($f_r$) of 0.25, i.e. for $A = 0$. An albedo of, say 0.1, further restricts the distribution 
to 0.2. Our results support similar conclusions arrived at by previous works (Deming et al. 2007; 
Spiegel et al. 2010). It is to be noted that we do not assume uniform illumination of the planetary 
dayside by the stellar irradiation (i.e weighing the stellar flux by $f = \frac{1}{2}$). Instead, we 
use $f = \frac{2}{3}$, according to the prescription of Burrows et al.~(2008), which is also used in 
Madhusudhan \& Seager (2009). Consequently, our estimation of T$_{\rm eff}$ for a given $\eta$ 
is typically higher than one would obtain using the $f = \frac{1}{2}$ assumption. 

Our conclusion of a low day-night redistribution on this planet assumes special significance 
for potential future observations of thermal phase curves  of GJ~436b. Thermal phase curves 
in the 3.6 $\micron$ and 4.5 $\micron$ IRAC channels, feasible with warm {\it Spitzer}, should 
show clear model-independent evidence of a high day-night temperature contrast, according to 
our present results.  A finding on the contrary, i.e finding efficient redistribution in the phase curves, 
can imply the possibility of a substantial interior energy source in GJ~436b. 

\section{Discussion and Summary}
\label{sec:discussion}

We presented a detailed analysis of the dayside atmosphere of GJ~436b. Our results show that 
a high metallicity along with non-equilibrium chemistry are required to explain the observations. 
We also studied the correlations between the various molecular species, and reported detailed 
constraints on the metallicity, chemical processes, and day-night energy circulation. 
Although our results come from observations in six channels of {\it Spitzer} photometry, some 
channels are more important than others. Here, we discuss the relative importance of the different 
{\it Sptizer} channels to our conclusions. We also discuss some potential alternate  
interpretations of the data, followed by a summary of our results. 

\subsection{Sensitivity of Results to {\it Spitzer} Observations}

The constraints reported in this work depend critically on the two {\it Spitzer} IRAC observations 
at 3.6 $\micron$ and 4.5 $\micron$. The high planet-star flux contrast in the 3.6 $\micron$ channel 
is responsible for the constraints of low methane abundance and low energy circulation.  The low 
flux contrast in the 4.5 $\micron$ channel is responsible for the requirement of high CO and/or 
CO$_2$ in GJ~436b. While the observation in the 3.6 $\micron$ channel was reported to be of 
the higher S/N of all channels, the 4.5 $\micron$ channel was a non-detection (Stevenson et al. 2010). 
Nevertheless, future observations in both these channels would be extremely important to confirm 
that these fluxes actually represent the steady state atmosphere in GJ~436b. The observations in 
the remaining four channels (5.8 $\micron$  - 24 $\micron$) are much less constraining, although 
still very useful. For instance, the moderate flux observed in the 8 $\micron$  channel, 
where methane absorbs strongly, is important to the conclusion that the very high 
flux in the 3.6 $\micron$ channel could not have been due to a thermal inversion causing 
methane emission. 

Our constraints on the molecular abundances are fairly robust with respect to the observational 
uncertainties. The high flux in the 3.6 $\micron$ channel cannot be explained by equilibrium 
chemistry. A model with equilibrium chemistry and solar or 30 $\times$ solar metallicities predicts 
planet-star flux contrasts that are lower than the observed value by over 4 -$\sigma$ (also in 
agreement with models of Demory et al. 2007; Spiegel et al. 2010). On the other hand, the contrasts 
predicted in the 4.5 $\micron$ channel based on equilibrium chemistry alone would be higher than 
the observed non-detection by over 3-$\sigma$ (also see Spiegel et al. 20010). We have been 
conservative in our analysis by allowing our best fits to the 4.5 $\micron$ point to lie within the 
3-$\sigma$ $\pm$ 1-$\sigma$ upper-limits. Had we considered this point to be a strict non-detection 
at 1-$\sigma$, our results would predict even higher CO and/or CO$_2$. Finally, the large 
uncertainties in the observed fluxes in the 16 $\micron$ and 24 $\micron$ channels provide only 
fiducial constraints on the temperature structure and the H$_2$O and CO$_2$ abundances. 

\subsection{Alternate Interpretations} 

The high planet-star flux contrast observed in the 3.6 $\micron$ IRAC channel is central 
to most of the constraints reported in this work. Our inferences could partly be restricted by 
the specific choices that are inherent to our models. For example, our models do not include 
clouds or hazes. Although scattering from hazes has been suggested to be potentially 
relevant in the optical and near-IR (e.g. Sing et al. 2009), a high contribution at the longer 
wavelengths of the 3.6 $\micron$ channel is unlikely. We have also assumed the planet 
atmosphere to be in local thermodynamic equilibrium (LTE). Swain et al. (2010) reported 
a potential signature due to non-LTE methane fluorescent emission in HD~189733b observed 
in the range of the 3.6 $\micron$ channel (but c.f. Mandell et al. 2010, who did not find such a 
feature in follow-up observations). The dayside atmosphere of GJ~436b might also be variable, 
as has been reported previously for hot Jupiters (Grillmair et al. 2008; Madhusudhan \& Seager, 
2009). However, for variability to explain the observed flux in the 3.6 $\micron$ channel, the 
temperature at the $\sim$ 1 bar level in the atmosphere of GJ~436b has to exhibit fluctuations 
greater than 400 K, between subsequent observations of Stevenson et al. (2010). These and 
other alternate explanations are worth exploring with more data at different epochs. 

\subsection{Summary}

We have presented constraints on the chemical composition and temperature structure 
of the dayside atmosphere of hot Neptune GJ~436b, based on recent {\it Spitzer} observations. 
One of our key findings is the strict upper limit on the mixing ratio of methane. We find that 
models fitting the observations require a methane mixing ratio below $10^{-6}$. Slightly higher 
methane mixing ratios require CO$_2$ $\sim 10^{-3} - 10^{-2}$, which is implausible 
in the hydrogen rich atmosphere with the temperature structure of GJ~436b. The abundances 
of all the molecules are highly correlated. Applying nominal conditions of physical plausibility, we 
find the constraints on the molecular mixing ratios to be CH$_4$ $ \sim 10^{-7} -  10^{-6}$, 
CO $\geq 10^{-3}$, CO$_2$ $\sim 10^{-6} - 10^{-4}$, and H$_2$O $\leq 10^{-4}$. 
These constraints on the molecular abundances cannot all be explained based on equilibrium 
chemistry, for any metallicity, as reported in Stevenson et al. (2010). At the temperatures of GJ~436b, 
equilibrium chemistry with solar abundances predicts CH$_4$, CO and CO$_2$ mixing ratios to 
be $\sim 5 \times 10^{-4}$, $\sim 10^{-5}$ and $\sim 10^{-7}$, respectively, contrary to the observed 
abundances.  

The observed constraints on the molecular abundances can be explained by a combination of 
high metallicity and non-equilibrium thermochemistry. A high metallicity is required for a high CO$_2$ 
abundance. Vertical mixing along with a high metallicity is required to dredge up the high CO abundance 
from the lower layers of the atmosphere to observable layers. Finally, vertical mixing and photochemistry can cause substantial depletion of CH$_4$, as reported by Zahnle et al. (2009b). At the temperatures (T $>$ 1100 K) we obtain for the lower atmosphere of GJ~436b, the results of Zahnle et al. (2009b) indicate the depletion of equilibrium CH$_4$ via oxidation to CO, caused by overabundance of the H radical. Our joint analysis of the parameters of non-equilibrium chemistry required to explain the abundances of all the species, suggests that the dayside atmosphere of GJ~436b has a high metallicity of $\sim 10 \times$ solar 
and a diffusion coefficient of $K_{zz} = 10^6 - 10^7$ cm$^2$/s. The metallicity is 
substantially enhanced over that of the host star which is consistent with 
solar metallicity (Torres et al. 2008). 

Our results also constrain the dayside temperature structure and the day-night energy 
redistribution in the atmosphere of GJ~436b. A temperature inversion is ruled out by the 
current observations; although, small inversions which are not observable at the resolution 
of the current photometric data cannot be conclusively ruled out. The observations also 
suggest inefficient day-night energy redistribution ($f_r$) in GJ~436b, requiring 
$(1-A_B)(1-f_r) = \eta \geq 0.7$, at the 1-$\sigma$ fit. Thus, the maximum $f_r$ allowed by 
the data at the 1-$\sigma$ fits is 0.3 for zero bond albedo ($A_B$), and 0.23 for $A_B$ = 0.1. 
Future observations of thermal phase curves in the available warm {\it Spitzer} channels will be 
instrumental in validating the low redistribution requirement. A finding on the contrary might 
indicate a substantial interior energy source. We emphasize that the constraints reported in 
this work depend primarily on the two {\it Spitzer} channels (3.6 $\micron$ and 4.5 $\micron$). 
Thus, future observations in these channels will be extremely important in confirming the present 
and previous results on the atmosphere of GJ~436b. 

The atmosphere of the hot-Neptune GJ~436b presents new challenges and opportunities 
for detailed modeling of exoplanet atmospheres. As low-mass transiting planets continue to be 
discovered, unexpected findings are likely to continue. The next generation of models and 
observations will help unravel those mysteries, and help put our solar system in perspective. 

\acknowledgements{NM thanks Adam Showman, James Kasting, Julianne Moses, Channon Visscher, David Spiegel, Jonathan Fortney, Ivan Hubeny, Olivier Mousis, and Mark Marley, for helpful discussions. We thank Richard Freedman for providing molecular line lists, especially with helpful information on CO$_2$ opacities. We thank Larry Rothman for access to the HITEMP database. We thank the MIT Kavli Institute for facilitating access to the computer cluster used for this work. Support for this work was provided by NASA through an award issued by JPL/Caltech.}

\end{document}